\documentclass[colorlinks]{article}
\usepackage[english]{babel}

\usepackage[nonatbib, preprint]{neurips_2024}
\usepackage[utf8]{inputenc}
\usepackage{biblatex}
\usepackage[utf8]{inputenc} 
\usepackage[T1]{fontenc}    
\usepackage{hyperref}       
\usepackage{url}            
\usepackage{booktabs}       
\usepackage{multirow}
\usepackage{nicefrac}       
\usepackage{microtype}      

\usepackage{subcaption}
\usepackage[autostyle, english = american]{csquotes}
\MakeOuterQuote{"}
\usepackage{anyfontsize}
\usepackage{stfloats}
\usepackage{algorithm2e}
\newcommand\mycommfont[1]

\usepackage{amsmath,amssymb,amsfonts}
\usepackage{cleveref}
\usepackage{algorithmic}
\usepackage{graphicx}
\usepackage{textcomp}
\usepackage{xcolor}
\def\BibTeX{{\rm B\kern-.05em{\sc i\kern-.025em b}\kern-.08em
    T\kern-.1667em\lower.7ex\hbox{E}\kern-.125emX}}

\title{Multi-Modality Transformer for E-Commerce: Inferring User Purchase Intention to Bridge the Query-Product Gap}

\author{%
  Srivatsa Mallapragada \\
  School of Data Science and Analytics\\
  Kennesaw State University\\
  Marietta, GA, USA\\
  \texttt{smallapr@gmail.com} \\
  \And
  Ying Xie\\
  Department of Information Technology \\
  Kennesaw State University\\
  Marietta, GA, USA\\
  \texttt{yxie2@kennesaw.edu} \\
  \And
  Varsha Rani Chawan \\
  The Home Depot, Inc \\
  Atlanta, GA, USA \\
  \texttt{varsha\_rani\_chawan@homedepot.com} \\
  \AND
  Zeyad Hailat \\
  The Home Depot, Inc \\
  Atlanta, GA, USA \\
  \texttt{zeyad\_hailat@homedepot.com} \\
  \And
  Yuanbo Wang \\
  The Home Depot, Inc \\
  Atlanta, GA, USA \\
  \texttt{cody\_wang1@homedepot.com} \\
}

\begin{document}

\maketitle

\begin{abstract}
  E-commerce click-stream data and product catalogs offer critical user behavior insights and product knowledge. This paper propose a multi-modal transformer termed as PINCER, that leverages the above data sources to transform initial user queries into pseudo-product representations. By tapping into these external data sources, our model can infer users' potential purchase intent from their limited queries and capture query relevant product features. We demonstrate our model's superior performance over state-of-the-art alternatives on e-commerce online retrieval in both controlled and real-world experiments. Our ablation studies confirm that the proposed transformer architecture and integrated learning strategies enable the mining of key data sources to infer purchase intent, extract product features, and enhance the transformation pipeline from queries to more accurate pseudo-product representations.
\end{abstract}

\section{Introduction}
\label{sec:intro}
E-commerce platforms generate vast amounts of click-stream data capturing users' shopping journeys. This data encompasses users' product searches, page clicks, cart additions, and purchases. When a user searches for a product, they typically click on various retrieval results before adding desired items to their cart. Analyzing these online shopping patterns provides insight into purchase intent - connecting queries to product clicks and cart adds. Additionally, aggregating data across users reveal diversity in product choice and purchasing behavior. While different users may search the same query, their subsequent clicks and purchases may widely vary.

\begin{figure}[ht]
\centering
\begin{subfigure}{0.5\linewidth}
    \centering
    \includegraphics[width=\linewidth, trim={0cm 0cm 10cm 0cm}, clip]{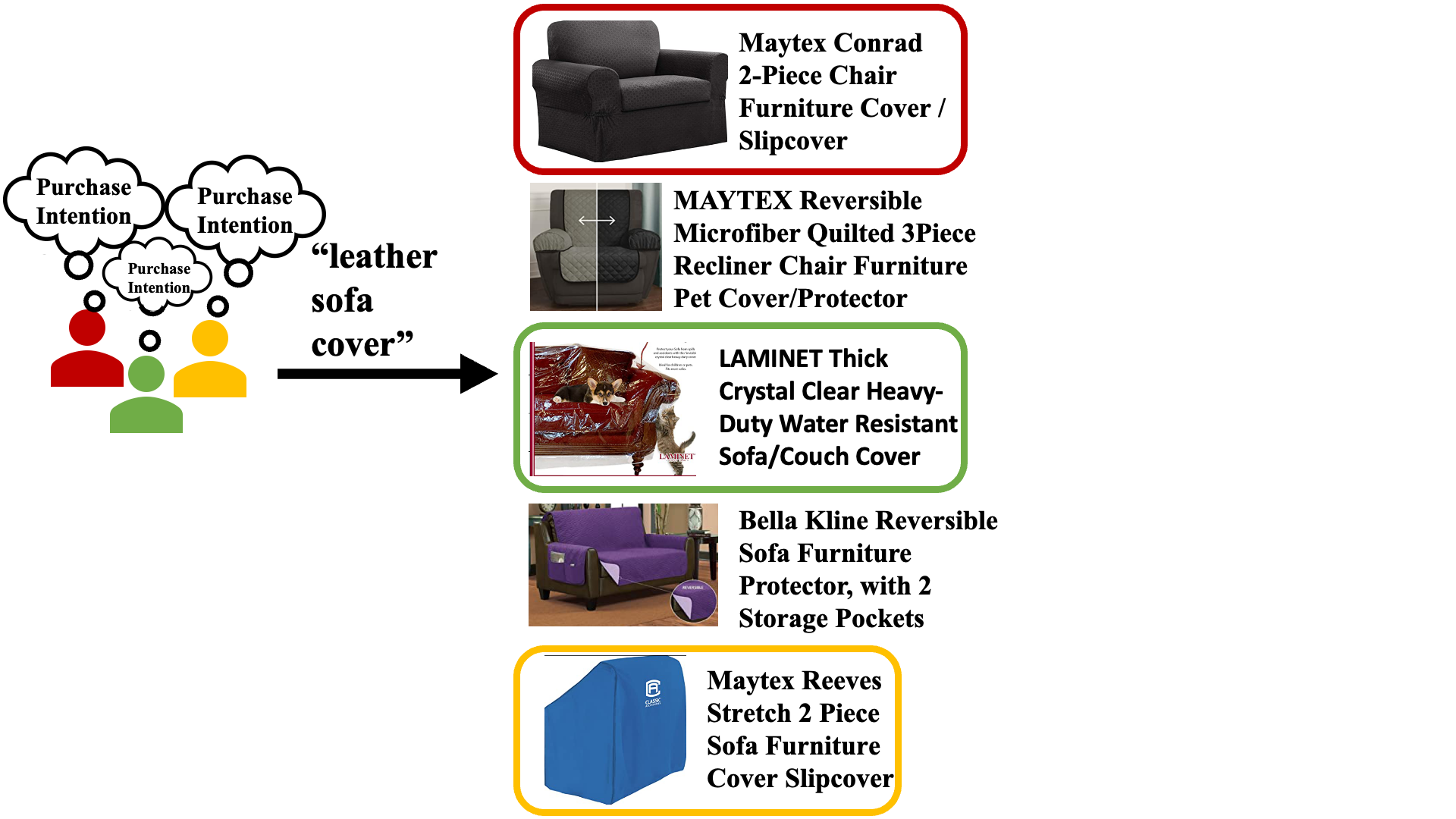}
    \caption{Purchase intention}
    \label{fig:am_example}
\end{subfigure}%
\begin{subfigure}{0.5\linewidth}
    \centering
    \includegraphics[width=\linewidth, trim={0cm 0cm 10cm 0cm}, clip]{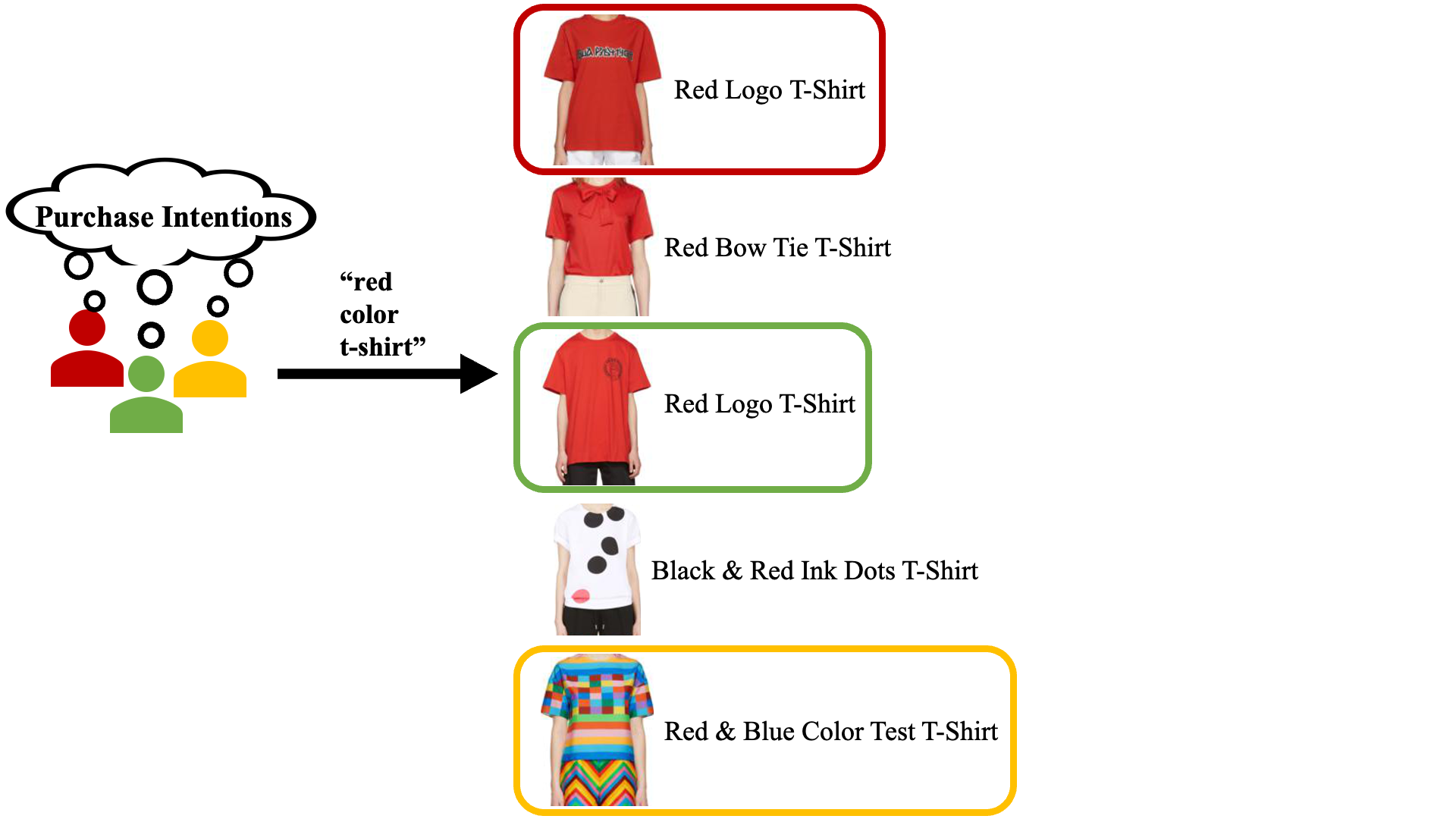}
    \caption{Product features}
    \label{fig:fg_example}
\end{subfigure}
\caption{Examples to show the importance of purchase intention and product features in users' purchase choices from a ranked list of products}
\vspace{-15pt}
\end{figure}

\subsection{Motivation 1: Purchase intention}
Click-stream data reveals that users with the same query may have different purchase intentions. Fig. \ref{fig:am_example} shows three users searching for a \textit{"\textbf{leather sofa cover},"} purchase different products. Current retrieval systems employ various techniques based on click-stream data to enhance query understanding and capture purchase intentions \cite{azad2019query, carmel2020multi, dai2023contrastive, guo2022semantic, zhou2021pssl}. However, these methods heavily depend on user queries and do not consider the purchased products associated with the query, creating a gap between the query and the purchased product. To bridge this gap, we utilize untapped information from query and Add To Cart (ATC) product pairs in click-stream data to transform the initial query into a pseudo product representation. This bridge can be defined as a purchase intention that is an amalgamation of information from query-ATC product pairs. To handle patterns of similar purchase intentions among users, we developed a reward-based competitive learning system that employs vector quantization to quantize crowd wisdom from query-ATC product pairs, inspired by Likas's work \cite{likas1999reinforcement}. After training, purchase intention embeddings provide supplementary information about the history of similar query-product pairs to the retrieval system, helping generate a new pseudo product embedding closer to the user's choice of potentially relevant products.

\subsection{Motivation 2: Granular Product Features}
Fig. \ref{fig:fg_example} illustrates a scenario where users exhibit similar purchase intentions, querying \textit{"\textbf{t-shirt with red color}"}, but they purchase different products. The product title purchased by the red, green, and yellow users contain \textit{"red"} and \textit{"t-shirt"} from the query, but the granular product text and image features contribute to users' preferences. A retrieval system should match the query at a finer granularity level to diversify the relevant products, aligning query words or sub-words with product words, sub-words, image patches, or kernels. Product features enable a retrieval system to focus on specific user preferences beyond their purchase intentions. This advocates for a system capable of processing both text and image product features from the catalog to meet users' needs. We consider this as a second motivation for our contributions and leverage multi-modal information from the product catalog, along with purchase intentions, to transform sparse queries into robust pseudo product representations for diverse, intent-aligned search.

\subsection{Proposed Multi-modal Transformer Framework}
The examples in figures \ref{fig:am_example} and \ref{fig:fg_example} highlight the limitations of current retrieval systems \cite{tsagkias2021challenges, chen2021towards, guo2022semantic} in capturing complex user purchase intents and retrieving diverse, intent-aligned products. To address this, we propose Purchase Intention-based Neural Causal E-commerce Retrieval (PINCER), a novel multi-modal transformer framework that transforms sparse queries into robust pseudo product representations by integrating extracted purchase intent vectors into the query transformation pipeline. PINCER is the first framework to connect queries and products through derived intent. A sequential two-stage training process estimates purchase intention and granular product features, from which the pseudo-product representation is generated. In stage 1, purchase intent vectors instigate shared learning between queries and products with reward based competitive learning, while granular product features are extracted from the product catalog and stored in a vector database. In stage 2, the decoder combines the trained purchase intent, retrieved product features, and query embedding to generate the pseudo product embedding, which is trained with preference modeling. This unified model transforms queries into pseudo product embeddings in real-time during inference. We conduct experiments on real-world home decor click-stream data and synthetic e-commerce datasets built from Amazon Cross-Market \cite{bonab2021crossmarket} and FashionGen \cite{rostamzadeh2018fashion} with non-linear image-based functions emulating as user purchase intentions to showcase PINCER's performance and learning capability.

The motivations outlined above propel our contributions, succinctly summarized as follows:
\begin{enumerate}
    \item[\textbf{1.}] We pioneer a multi-modal transformer framework, uniquely integrating purchase intention vectors and multi-modality granular product features into an e-commerce retrieval.
    \item[\textbf{2.}] We employed a reward-based competitive learning to extract purchase intention from the click-stream data, and adapted preference modeling to generate the pseudo product embedding with purchase intention based negative sampling. 
    \item[\textbf{3.}] Leveraging PINCER, we demonstrate the superior performance of our model, surpassing baseline multi-modal and text modality e-commerce retrieval models by an impressive 10.81\% (Recall) on the real-world e-commerce experiments. 
\end{enumerate}

\section{Related Work}
PINCER is a novel multi-modal transformer framework, designed to transform a query text input into a pseudo product embedding and retrieve most relevant products from the product catalog. It accommodates relevance ranking at it's core, but distinct from existing e-commerce product search systems \cite{van2016learning, ai2017learning, guo2018multi, nigam2019semantic, zhang2020towards, li2021embedding, zheng2023make}. PINCER advances beyond relevance ranking by extracting purchase intents from the query transaction history, and generate new pseudo product embeddings for product retrieval.

\subsection{Purchase intention estimation}
PINCER defines purchase intention as a user's potential intention to add a specific product to the shopping cart, associating a query with the ATC product. Although PINCER shares similarities with personalized search systems \cite{lee2005automatic, dai2006detecting, guo2010ready, lo2016understanding, ai2017learning, guo2019attentive, wu2019influence, zhang2020towards, bi2020transformer, hendriksen2020analyzing, zhou2021pssl, dai2023contrastive}, it is not one. Personalized search systems incorporate user profile information, queries, and product selections for user-specific search, while PINCER improves core product search with a novel data extraction and pseudo product embedding generation pipeline. Personalized search can filter and rank products tailored to users but suffers from cold start problems without prior user history \cite{bennett2012modeling} and only benefits high-entropy queries \cite{dou2007large}. In contrast, PINCER does not have cold start issues and can be used on any query from any user. PINCER's purchase intention vectors represent crowd wisdom linking queries to products, not individual user preferences. Unlike personalized search systems utilizing user information from click-stream data, intention vectors are estimated from query-product pairs. It involves incremental latent space learning, manifesting as shared vector representations useful as centroids for similar pairs. Although existing literature lacks a direct implementation of these vectors, we designed a competitive learning training strategy with vector quantization. While SwAV \cite{caron2020unsupervised} and MoCo \cite{he2020momentum} show self-supervised online clustering methods that may suffice our needs, they fall short for intent estimation due to disjoint query and product data sources. Critically, they do not enforce pairing queries and products to the same cluster center during training, which is essential for modeling intent via purchase history. Some retrieval studies \cite{zhang2016collaborative, yu2018product, jang2021self} demonstrate query-product space quantization but share similar limitations by not aligning query-product pairs to the same intent vector. Inspired by Likas \cite{likas1999reinforcement}, we view query-product pairs alignment as a Bernoulli trial and use reward-based competitive learning with vector quantization to enforce each pair to select the same purchase intention vector and quantize the latent space. This quantization enables PINCER to model purchase intents as vectors capturing associations between queries and products.

\subsection{Granular product features in retrieval}
PINCER, during training, aligns both text and image granular product features with the query features. It later stores individual text and image features from all the catalog products in vector databases. PINCER, during training and inference, retrieve these query aligned product features for pseudo product generation. Existing approaches like Pseudo Relevance Feedback (PRF) \cite{wang2021pseudo, yu2021improving} use product information to expand the query context and perform a secondary retrieval. This uses a two stage retrieval approach that is relied on overall product information to improve query embeddings. Unlike PRF using individual product embeddings, PINCER aims to utilize the individual product features aligned with query tokens to generate a new pseudo product embedding. This requires the storage of granular features for a real-time retrieval during the query transformation process. This makes PINCER unique to store and retrieve product features when required. 

\subsection{Multi-modal transformer for retrieval}
PINCER, as a multi-modal framework, transforms a textual query into a pseudo product embedding aligned with product embeddings from the catalog. Stage 1 training semantically maximizes the similarity between query and product embeddings in the shared latent space for further transformation, employing contrastive learning strategies that have demonstrated success in e-commerce product and personalized search models \cite{guo2018multi, izacard2021unsupervised, zhou2021pssl, ma2022ei, hendriksen2022extending, shin2022clip, dai2023contrastive}. Research in text-to-image ($t2i$) retrieval, such as ViLBERT \cite{lu2019vilbert}, UNITER \cite{yen2020uniter}, ALIGN \cite{li2021align}, ViLT \cite{kim2021vilt}, and BLIP \cite{li2022blip}, highlights the learning of feature extraction and fusion of modalities to derive comprehensive representations. PINCER employ pre-trained encoders for query and product encoding in stage 1. This methodology uses a two-tower architecture \cite{yu2022commercemm, liu2022universal, wang2023one} to accommodate query text alignment with product text and image embeddings using the contrastive learning strategy. Stage 2 training combines purchase intention, product features, and query within the latent space to generate a composite pseudo product embedding, leveraging causal attention in a transformer decoder. The decoder adapts preference modeling as a training strategy, utilizing soft positive and negative rewards. Amanda et al. \cite{askell2021general} showed a use case of preference modeling pre-training, utilizing a binary reward system to enhance sample efficiency in Large Language Models (LLMs). This encouraged us to choose a positive target product and sample negative targets from a neighborhood of products around the target product purchase intention vector for training. This makes the decoder generate embeddings closer to the target product within the neighborhood of similar products. 

\section{Methodology}
Let $D = (X,Y)$ denote the observable query-product pairs from the click-stream data, where $X = \{x_{1}, x_{2}, x_{3}, \dots, x_{N}\}$ represents the set of user queries and $Y = \{y_{1}, y_{2}, y_{3}, \dots, y_{N}\}$ represents the set of corresponding products added to the cart. Each query $x_{n}$, governed by the vocabulary of the query text encoder, may vary in length, comprising (sub)words. Similarly, the product set $Y$ follows suit, defined by the vocabulary of the product text encoder. In the case of image modality products, each $y_{n}$ consists of a sequence of fixed-length image patches, as per the constraints of the image encoder. Here, $N = |D|$ denotes the size of the training data, and $(x_{n}, y_{n}) \in \mathcal{R}^{d}$, residing in a shared latent embedding space of dimensionality $d$. We select $d = 128 | 256$ to optimize computational efficiency during retrieval. Importantly, it's worth noting that $D$ encompass repetitions of $x_{n}$, $y_{n}$, but no $(x_{n}, y_{n})$ pairs. A set of users' purchase intentions are denoted as $S = \{s_{1}, s_{2}, s_{3}, \dots\ s_{K}\}$ with $s\ \epsilon\ S$ reflecting a distinct intention marginalized over queries and purchased products. A set of product text and image features derived from catalog products are represented as $F=\{f_{1},f_{2},f_{3}, \dots f_{J}\}$ with $f\ \epsilon\ F$, and $f_{j}=(f_{j_{t}}, f_{j_{i}})$ being an ordered set of text and image representations.

\subsection{Model Architecture}
\begin{figure*}[!h]
  \centering 
  \includegraphics[width=\linewidth]{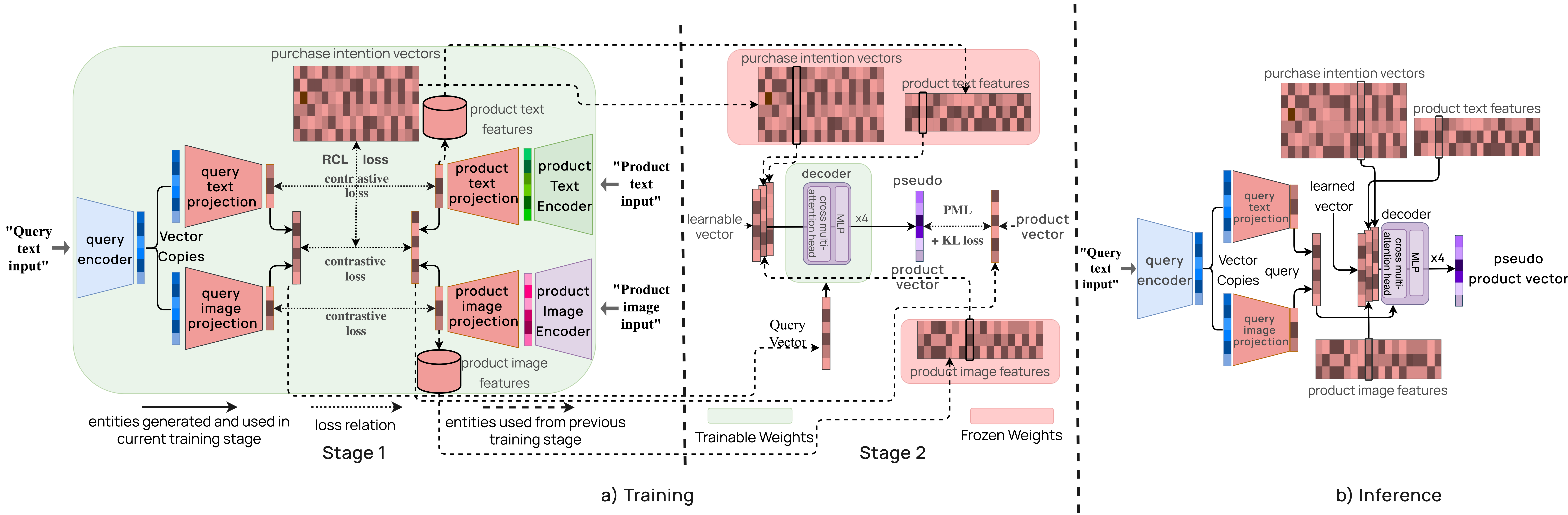} 
  \caption{PINCER model architecture in various stages of training} 
  \label{fig:PINCER_arch} 
\end{figure*}
PINCER is a unified model that improves recall over precision to allot more relevant and diverse products in a ranked list of relevant products. PINCER maximize respective probabilities $P(y_{n}|x_{n})$, $P(x_{n}|y_{n})$, $P(s_{k}|x_{n})$, $P(s_{k}|y_{n})$, and $P(f_{j}|x_{n})$ in stage 1 training using query $x_{n}$ and product $y_{n}$ pairs. $P(y_{n}|x_{n})$, and $P(x_{n}|y_{n})$ aligns the query and product embeddings \cite{radford2021learning}, whereas $P(f_{j}|x_{n})$ aligns the product features with the query token embeddings. The reward-based competitive training of $s_{k}$ maximizes $P(s_{k}|x_{n})$ and $P(s_{k}|y_{n})$ by reducing the distance between $x_{n}$ and $s_{k}$, and $y_{n}$ and $s_{k}$. This joint training enhances query-product alignment for retrieval and intent estimation. Stage 2 training aims to maximize the conditional probability $P(y_{n}|x_{n}, s_{k}, f_{j})$ by conditioning the query $x_{n}$ on product features $f_{j}$ and purchase intention vector $s_{k}$ chosen by $x_{n}$ to generate pseudo product $\widehat{y_{n}}$ close to target $y_{n}$. 

\subsection{\textbf{Stage 1: Purchase intention estimation and relevance ranking}}
The concept of purchase intention is central to understanding the methodology of the PINCER algorithm. In e-commerce, users typically have a purchase intention in mind when they search a product. This intention is often reflected in their search query and thereby their choice of 
the products added to their cart. By capturing and modeling these purchase intentions, PINCER aims to bridge the gap between user queries and relevant products, improving the overall retrieval process. The purchase intention vectors in PINCER serve as a representation of the latent space where queries and products are associated based on historical user behavior. These vectors act as a crowd purchase intentions for aligning diverse yet similar purchase intentions of users. PINCER understand user preferences from the crowd purchase intention vectors and combines the information with incoming query to retrieve products that closely match users intentions.

PINCER, depicted in Fig \ref{fig:PINCER_arch}a), incorporates encoders and a pseudo product decoder trained across stages. The query embeddings, derived from the encoders, are projected into both product text and image spaces. This process aims to optimize the selection of the most relevant product feature for a given user query. The resultant query embedding, formed by concatenating these projections, aligns with the product embedding concatenated from product text and image embeddings. These semantic training strategies employ a contrastive loss from (\ref{eq:contrastive_loss}), utilizing parameters such as batch size ($B$) and temperature ($\mathcal{T}$), akin to the approach in CLIP \cite{radford2021learning}. This contrastive loss proves instrumental in semantically ranking embeddings for dense retrieval purposes \cite{zhou2021pssl, izacard2021unsupervised, ma2022ei}. The contrastive loss between query and product text encoders is represented as $L_{qpt}$, image encoders is $L_{qpi}$, and concatenated vectors is $L_{qp}$ from (\ref{eq:stage_1_loss}). 

\begin{equation}
    L_{cont} = - \sum_{n=1}^{N}\log{\frac{exp(x_{n}y_{n}^{\intercal}/\mathcal{T})}{\sum_{b=1}^{B}exp(x_{n}y_{b}^{\intercal}/\mathcal{T})}}, 
    \label{eq:contrastive_loss}
\end{equation}

The estimation of purchase intention begins by initializing a uniformly distributed fixed set of vectors that match the dimensions of concatenated query and product vectors. Inspired by competitive learning \cite{likas1999reinforcement}, queries and products select the nearest intent vector by Euclidean distance. Assuming user's intention is the latent link between a query and ATC product, a reward system (\ref{eq:pur_int_rll}) positively rewards the query-product pair to choose the same purchase intention and pushes the intent vector towards the query-product pair. Mismatched choices get negative rewards, separating query-product-intent vectors. The remaining probability $rp_{@_{n},s_{k_{@}}}$ from (\ref{eq:reward}) serves as a learning rate, balancing loss updates between converging and diverging vectors. Since the choice of closest intent vector for query or product is a binary operation, the selection probability (\ref{eq:prob}) converts query/product-intent distances into Bernoulli probabilities. This lowers the remaining probability as distances decrease, controlling the loss for tuning query, product, and the purchase intention vectors.
\begin{equation} 
    L_{stage_1} = \lambda(L_{qp})+ (1-\lambda)(L_{qpt} + L_{qpi}) + RCL
  \label{eq:stage_1_loss} 
\end{equation}
\begin{equation}
    RCL = 0.5*(r_{s_{k}}*rp_{x_{n},s_{k_{x}}}*||x_{n} - s_{k_{x}}||_{2} \\
    + r_{s_{k}}*rp_{y_{n},s_{k_{y}}}*||y_{n} - s_{k_{y}}||_{2})
    \label{eq:pur_int_rll}
\end{equation}
\begin{equation}
    s_{k_{@}} = argmin_{S}(||@_{n} - S||_{2}),\ where\ @=x\ or\ y
    \label{eq:nearest_intent}
\end{equation}
\begin{equation}
    r_{s_k} = \begin{cases}
                1, & \text{if $s_{k_{x}} = s_{k_{y}}$}\\
                -1, & \text{otherwise}.
              \end{cases}
    \label{eq:reward}
\end{equation}
\begin{equation}
    p_{@_{n}, s_{k_{@}}} = 2*\big(1 - \frac{1}{1+\exp^{-||@_{n} - s_{k_{@}}||_{2}}}\big), \\
    \ where\ @=x\ or\ y
    \label{eq:prob}    
\end{equation}
\begin{equation}
    rp_{@_{n}, s_{k_{@}}} = \begin{cases}
                                1-p_{@_{n}, s_{k_{@}}},&\text{if $s_{k_{x}} = s_{k_{y}}$}\\
                                -p_{@_{n}, s_{k_{@}}}, & \text{otherwise}.
                           \end{cases}\\
    ,\ where\ @=x\ or\ y
    \label{eq:remaining_prob}
\end{equation}

The Stage 1 loss, in (\ref{eq:stage_1_loss}), combines contrastive losses and Reward based Competitive Learning (RCL) loss for purchase intention. The parameter $\lambda$ balances query-product embedding similarity, with an optimal setting at $\lambda = 0.5$ for effective learning in both encoders. Embeddings are normalized to a range of $[-1,1]$, enhancing stability and facilitating the use of dot product for product retrieval.\\

\subsection{\textbf{Stage 2: Pseudo product embedding generation}}
The generation of pseudo product embeddings in PINCER is crucial for bridging the gap between user queries and relevant products. By combining purchase intention, product features, and query information, PINCER generates a comprehensive embedding capturing user preferences and essential product characteristics. This generated embedding serves as  a proxy for the ideal product that the user may potentially purchase.

In the stage 2 training (Fig \ref{fig:PINCER_arch}b), a transformer decoder with causal attention mask sequentially combines purchase intention embedding, product feature embeddings, and a learnable bias vector with cross-attending query vector to generate the pseudo product embedding. The parameters of encoders and intention vectors are frozen during the training phase. Product text and image feature embeddings (tokenized titles and images) are stored in a faiss \cite{johnson2019billion} vector database for real time retrieval. Each query retrieves the most relevant product feature vectors via vector search. The retrieved text and image vectors are concatenated along query token positions, matching the decoder's working dimensions. The query vector provides context to the decoder to generate the embedding closer to the target product embedding.
\begin{equation} 
  L_{stage_2} =  PML + KL(SD_{\widehat{y_{n}}, y_{n}}||SD_{x_{n}, y_{n}}) 
  \label{eq:stage_2_loss} 
\end{equation}
\begin{equation}
  PML = \sum_{n=1}^{N}-log(\exp^{(\widehat{y_{n}}*y_{n} - \widehat{y_{n}}*y_{+})}),\ where\ y_{+}\neq y_{n}
  \label{eq:pref_reward} 
\end{equation}
\begin{equation}
    SD_{@_{n}, y_{n}} = log\Big(\frac{exp(@_{n}y_{n}^{\intercal})}{\sum_{b=1}^{B}exp(@_{n}y_{b}^{\intercal})}\Big)
    \label{eq:sim}
\end{equation}
The decoder is trained to increase the precision within the target product neighborhood while retaining the stage 1 recall from query-product relevance ranking (\ref{eq:stage_2_loss}). The training process involves computing a soft reward based on cosine similarity between pseudo product and target product embeddings called as Preference Modeling Loss (PML)(\ref{eq:pref_reward}). Negative samples $y_{n}$, sourced from nearby target product $y_{+}$, are efficiently retrieved from product clusters using the intention vectors as cluster centers. Employing Kullback–Leibler divergence helps align distribution of similarities between query-target products and pseudo-target products. This divergence shown in (\ref{eq:sim}) guides the decoder in improving the pseudo product generation and retaining the recall without overly separating negative samples. This combined loss function from (\ref{eq:stage_2_loss}) enhances the model to capture user preferences more effectively than standard e-commerce search systems. During Stage 1 training, multiple objectives are jointly optimized using a weighted sum to maximize the probabilities. Subsequently, Stage 2 training optimizes single objective with a decoder component attached to the Stage 1 trained checkpoint with frozen parameters, maximizing the overall retrieval performance.

\subsection{\textbf{Inference:}}
During inference (Fig. \ref{fig:PINCER_arch}), PINCER utilizes the query encoder to project input text into the shared latent space. The resulting query embedding selects the nearest purchase intent vector and product features from the database. The decoder takes the intent, features, and bias vectors as input, with query vector cross-attending over the input vectors, to output the final pseudo product embedding. PINCER leverages the query encoder and the decoder to generate pseudo product embeddings from input text and retrieve relevant products in real-time. All product embeddings are pre-generated from the catalog and stored externally. The generated pseudo-product embeddings are matched against pre-stored product embeddings using cosine similarity to retrieve the top-k matches. This similarity-based ranking enables efficient retrieval of relevant products in real-time. For optimization, we employ vector clustering on product vectors using the purchase intent vectors as cluster centers. This reduces the retrieval time by pre-selecting clusters via the purchase intention vector that aligns with incoming query.

\section{Experiments}
This study tests PINCER model on real world experiments with real-world data and controlled experiments to test the model's capability to learn the synthetic purchase intention modeled data.

\subsection{Real-world experiments:}
We evaluate PINCER using real-world e-commerce click-stream data from a large company, containing 1.38M training and 173K validation/testing query-product pairs across 212K products in 4 home decor categories. The data includes user queries and their corresponding add-to-cart products, capturing real purchase intents through pairs of searches and items added to carts. This tests PINCER's ability to deduce purchase intents from crowd patterns and leverage them to improve product search over state-of-the-art (SOTA) baselines. Experiments on these query-product pairs with direct user actions validate whether PINCER can effectively extract and apply purchase intents to advance retrieval. Overall, these real e-commerce interactions provide an authentic test-bed for evaluating our approach's capabilities in a production environment. 

\subsection{Controlled experiments:}
\begin{figure}[h]
    \centering
    \includegraphics[scale=0.23, trim={0 0cm 0 1cm}, clip]{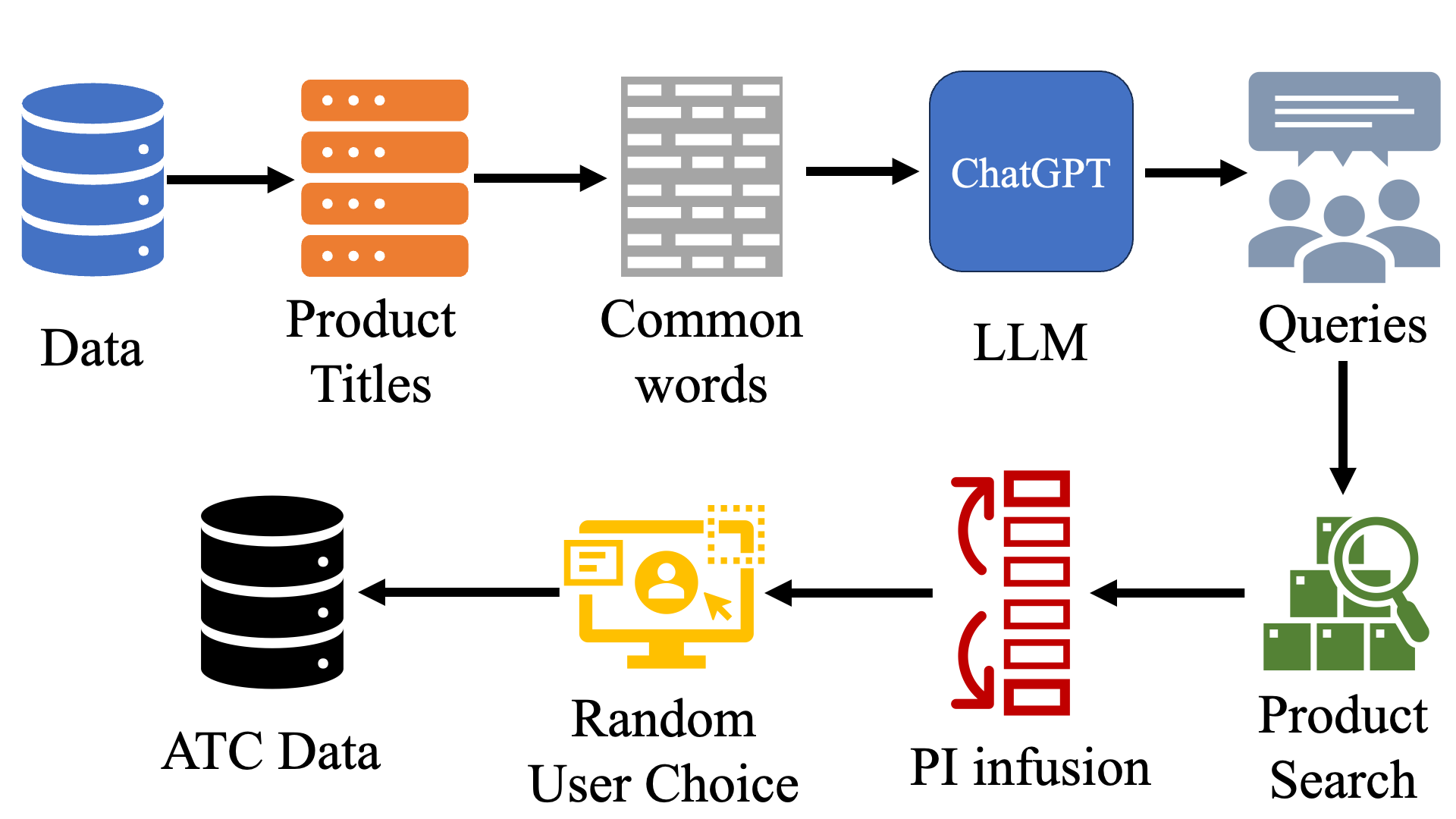}
    \caption{Synthetic Add-To-Cart (ATC) data generation from FashionGen and Amazon datasets using a LLM}
    \vspace{-10pt}
    \label{fig:data_gen}
\end{figure}
We conduct controlled experiments using synthetic datasets with queries generated for product catalogs from Amazon Cross-Market \cite{bonab2021crossmarket} (124K products, 44 categories) and FashionGen \cite{rostamzadeh2018fashion} (67K products, 48 categories) via ChatGPT $3.5$ (LLM). The purpose is to embed synthetic purchase intentions to mimic click-stream transaction logs. Considering the Fig. \ref{fig:data_gen}, products were grouped by taxonomy and titles clustered to extract top words for query creation. ChatGPT $3.5$ produced $5$ customer-like queries per group. Products were ranked via text based semantic retrieval and filtered by image brightness/gradient as a Purchase Intention (PI) infusion process to isolate simulated cart adds. The color brightness and product patterns may reflect user choices as purchase intention and specificity in granular features. $1-5$ random products per query were selected to simulate a search with one user-choice from 5 pages with $20$ products per page. Product-specific queries were also generated, resulting in 238K training, 29K validation/testing pairs for FashionGen, and 844K training, 105K validation/testing pairs for Amazon. The synthetic benchmarks enable evaluating the model with tailored search behaviors. These datasets are available for free download at \href{https://osf.io/qm65s/?view_only=c9dfd524c2f14e77a000300d4c71a0ad}{\textcolor{blue}{OSF}}.

\subsection{Implementation Details}
PINCER framework can accommodate light weight pre-trained encoders to LLMs that generate semantic embeddings. However, we decided to show the efficiency of the framework with light weight encoders competing with SOTA multi-modal retrieval systems. We employ distilBERT for text encoding and ResNet-50 for image encoding, both pre-trained. These models are projected to a $128-$dimensional space for efficiency. Each projector consists of a feedforward layer with GELU activation, post-activation layer norm, and $10\%$ dropout. AdamW optimizer with weight decay $1e-4$ is used, adjusting learning rate on plateau. Training spans 15 epochs for each stage, with epochs varying based on dataset size. Models are trained on a Quadro RTX 6000 GPU (24GB RAM) for synthetic datasets and an A100 40GB GPU for real-world data with 3 hours of training per epoch.

\subsection{Evaluation}
We assess our e-commerce retrieval framework by matching text queries with text and image products, using evaluation metrics such as precision and recall at various top values ($10, 20, 50,$ and $100$). We compare PINCER's recall using $SumR=(Recall@10+Recall@20+Recall@50+Recall@100)$ with some baselines because it is a  holistic measure of cumulative recall across top-k retrieval values, which aligns with similar retrieval benchmarks in e-commerce. This metric ensures a comprehensive view of the model's performance across various levels of user interaction. The chosen baselines—CLIP, FashionCLIP, and RetroMAE \cite{liu2022retromae}—are well-established models in large-scale e-commerce retrieval and have demonstrated strong performance in this domain. CLIP provides a general multi-modal benchmark, FashionCLIP is tailored to fashion retrieval tasks, and RetroMAE optimizes retrieval-oriented embeddings for e-commerce with a dual encoder approach for the downstream task. To ensure comparability, all baselines were fine-tuned on task-specific datasets under consistent evaluation settings. In our assessment, we use fully trained PINCER outcomes to exhibit its advancements over domain-specific, text, and multi-modality models. Other e-commerce models mentioned in the literature are neither openly available nor trainable in a reasonable amount of time with our GPU resources.


\vspace{-4pt}
\subsection{Quantitative Results}
Table \ref{tab:real_pre_rec} shows PINCER's superior precision and recall across all metrics, with significant improvements over other methods. PINCER achieved a $10.81\%$ boost in overall recall on real-world data compared to the closest baseline. Crucially, it had markedly higher recall at the top 10 and 20 products, which is critical for e-commerce as users are more likely to purchase from the first page of results. PINCER's substantial early recall improvements demonstrate its ability to reliably surface relevant products within immediate view, enhancing user experience and driving business metrics like engagement and conversion. The results prove PINCER's real-world value in dramatically improving top-ranked retrieval and customer experience.

\begin{table*}[ht]
    \fontsize{9pt}{9pt}\selectfont
    \centering
    \resizebox{\linewidth}{!}{%
    \begin{tabular}{@{}|c|c|p{1cm}p{1cm}p{1cm}p{1cm}|p{1cm}p{1cm}p{1cm}p{1cm}|p{1cm}|@{}}
    \toprule
    \multirow{2}{*}{Dataset} & \multirow{2}{*}{Model} & \multicolumn{4}{c|}{Precision} & \multicolumn{4}{c|}{Recall} & \multirow{2}{*}{SumR} \\ \cmidrule(lr){3-10}
                                & & P@10    & P@20   & P@50  & P@100  & R@10   & R@20   & R@50  & R@100  &  \\ \cmidrule{1-11}
    \multirow{4}{*}{Real world} & RetroMAE & 4.15   & 2.65  & 1.38  & 0.81  & 38.76  & 48.72  & 62.28  & 71.74  & 222 \\ \cmidrule{2-11}
                    & CLIP & 0.83  & 0.53 & 0.30 & 0.19 & 7.82 & 10.02 & 10.68 & 13.94 & 42.46 \\ \cmidrule{2-11}
                    & FashionCLIP & 0.94  & 0.61  & 0.34  & 0.22  & 8.87 & 11.46 & 15.91 & 20.08 & 56.32 \\ \cmidrule{2-11}
                    & \textbf{PINCER} & \textbf{4.94} & \textbf{3.08} & \textbf{1.54} & \textbf{0.87} & \textbf{45.47} & \textbf{55.74} & \textbf{68.26} & \textbf{76.4} & \textbf{246} \\ \cmidrule{2-11}
                    & $\uparrow$\ Relative & 19.03\%  & 16.22\% & 11.59\% & 7.41\% & 17.31\% & 14.41\% & 9.6\% & 6.49\%  &  10.81\%                  \\ \bottomrule
    \end{tabular}%
    }
    \caption{Comparison of retrieval models on real-world Add-To-Cart (ATC) transaction history. $\uparrow$ is a relative improvement with the RetroMAE(text model) metrics}
    \vspace{-10pt}
    \label{tab:real_pre_rec}
\end{table*}  

\begin{table*}[ht]
    \fontsize{9pt}{9pt}\selectfont
    \centering
    \resizebox{\linewidth}{!}{%
    \begin{tabular}{@{}|c|c|p{1cm}p{1cm}p{1cm}p{1cm}|p{1cm}p{1cm}p{1cm}p{1cm}|p{1cm}|@{}}
    \toprule
    \multirow{2}{*}{Dataset} & \multirow{2}{*}{Model} & \multicolumn{4}{c|}{Precision} & \multicolumn{4}{c|}{Recall} & \multirow{2}{*}{SumR} \\ \cmidrule(lr){3-10}
                                & & P@10    & P@20   & P@50  & P@100  & R@10   & R@20   & R@50  & R@100  &  \\ \cmidrule{1-11}
    \multirow{4}{*}{\begin{tabular}[c]{@{}c@{}}FashionGen \\ (brightness)\end{tabular}} & RetroMAE & 4.23 & 3.07  & 1.88  & 1.17  & 30.45  & 41.60  & 60.10  & 72.70  & 205 \\ \cmidrule{2-11}
                    & CLIP & 1.54  & 0.91  & 0.46  & 0.27  & 14.03 & 16.11 & 19.15 & 21.55 & 71 \\ \cmidrule{2-11}
                    & FashionCLIP & 1.75  & 1.03  & 0.50  & 0.29  & 15.85 & 18.05 & 20.63 & 22.89 & 77 \\ \cmidrule{2-11}
                    & \textbf{PINCER} & \textbf{4.73} & \textbf{3.56} & \textbf{2.13} & \textbf{1.27} & \textbf{31.18} & \textbf{45.35} & \textbf{65.94} & \textbf{78.07} & \textbf{221} \\ \cmidrule{2-11}
                    & $\uparrow$\ Relative & 11.82\%  & 15.96\% & 13.29\% & 8.55\% & 2.40\% & 9.01\% & 9.72\% & 7.38\% &  7.8\%                  \\ \midrule

    \multirow{4}{*}{\begin{tabular}[c]{@{}c@{}}FashionGen \\ (mean gradient)\end{tabular}} & RetroMAE & 4.28   & 3.09  & 1.88  & 1.17  & 30.66  & 41.78  & 60.15  & 72.80  & 205 \\ \cmidrule{2-11}
                    & CLIP & 1.49  & 0.88  & 0.44  & 0.26  & 13.61 & 15.67 & 18.51 & 20.97 & 69 \\ \cmidrule{2-11}
                    & FashionCLIP & 1.69  & 0.97  & 0.47  & 0.27  & 15.59 & 17.44 & 19.94 & 22.16 & 75 \\ \cmidrule{2-11}
                    & \textbf{PINCER} & \textbf{4.81} & \textbf{3.64} & \textbf{2.17} & \textbf{1.29} & \textbf{31.85} & \textbf{46.36} & \textbf{67.08} & \textbf{79.01} & \textbf{224} \\ \cmidrule{2-11}
                    & $\uparrow$\ Relative & 12.38\%  & 17.79\% & 15.43\% & 10.26\% & 3.88\% & 10.96\% & 11.52\% & 8.53\% &  9.27\%                  \\ \midrule
                    
    \multirow{4}{*}{\begin{tabular}[c]{@{}c@{}}Amazon \\ (brightness)\end{tabular}} & RetroMAE & 3.62   & 2.52  & 1.42  & 0.86  & 28.24  & 38.84  & 53.46  & 64.08  & 185 \\ \cmidrule{2-11}
                    & CLIP & 0.86  & 0.60  & 0.37  & 0.24  & 7.10 & 9.86 & 14.74 & 19.32 & 51 \\ \cmidrule{2-11}
                    & FashionCLIP & 0.97  & 0.68  & 0.41  & 0.27  & 8.01 & 11.03 & 16.43 & 21.36 & 57 \\ \cmidrule{2-11}
                    & \textbf{PINCER} & \textbf{4.54} & \textbf{3.15} & \textbf{1.68} & \textbf{0.96} & \textbf{34.55} & \textbf{47.56} & \textbf{62.65} & \textbf{71.59} & \textbf{216} \\ \cmidrule{2-11}
                    & $\uparrow$\ Relative & 25.41\%  & 25\% & 18.31\% & 11.63\% & 22.34\% & 22.45\% & 17.19\% & 11.72\% &  16.76\%                  \\ \midrule

    \multirow{4}{*}{\begin{tabular}[c]{@{}c@{}}Amazon \\ (mean gradient)\end{tabular}} & RetroMAE & 3.51 & 2.48  & 1.41  & 0.86  & 27.26  & 37.9  & 52.8  & 63.94  & 182 \\ \cmidrule{2-11}
                    & CLIP & 0.86  & 0.60  & 0.37  & 0.24  & 6.91 & 9.71 & 14.62 & 19.09 & 51 \\ \cmidrule{2-11}
                    & FashionCLIP & 0.97  & 0.68  & 0.41  & 0.27  & 8.11 & 11.10 & 16.37 & 21.38 & 57 \\ \cmidrule{2-11}
                    & \textbf{PINCER} & \textbf{4.55} & \textbf{3.14} & \textbf{1.69} & \textbf{0.97} & \textbf{34.52} & \textbf{47.29} & \textbf{62.89} & \textbf{72.09} & \textbf{217} \\ \cmidrule{2-11}
                    & $\uparrow$\ Relative & 29.34\%  & 26.61\% & 19.85\% & 12.79\% & 26.63\% & 24.77\% & 19.11\% & 12.74\% &  19.23\%                  \\ \bottomrule
    \end{tabular}%
    }
    \caption{Comparison of retrieval models on five datasets. $\uparrow$ is a relative improvement with the RetroMAE(text model) metrics}
    \vspace{-15pt}
    \label{tab:syn_pre_rec}
\end{table*}

Table \ref{tab:syn_pre_rec} shows PINCER outperformed RetroMAE, CLIP, and FashionCLIP on four synthetic datasets from FashionGen and Amazon, with $8.5\%-17.99\%$ average recall improvements over RetroMAE. Despite poorer overall performance, FashionCLIP slightly edged CLIP due to fashion-specialized training. The models' steady recall across datasets proves the benchmarks' reliability for validating specialized methods like PINCER. A Wilcoxon signed-rank test between RetroMAE and PINCER ($p-value = 0.043114$) at $90\%$ confidence level demonstrated PINCER's statistically significant performance gains over RetroMAE. By surpassing strong baselines on controlled synthetic data, PINCER displays robust improvements independent of real-world biases, reinforcing its strengths in aligning searches and products through pseudo product representations.

\subsection{Qualitative Results}
\begin{figure*}[!h]
\centering
\begin{subfigure}{0.49\linewidth}
  \centering
  \includegraphics[width=\linewidth, trim={0 0 0 0}, clip]{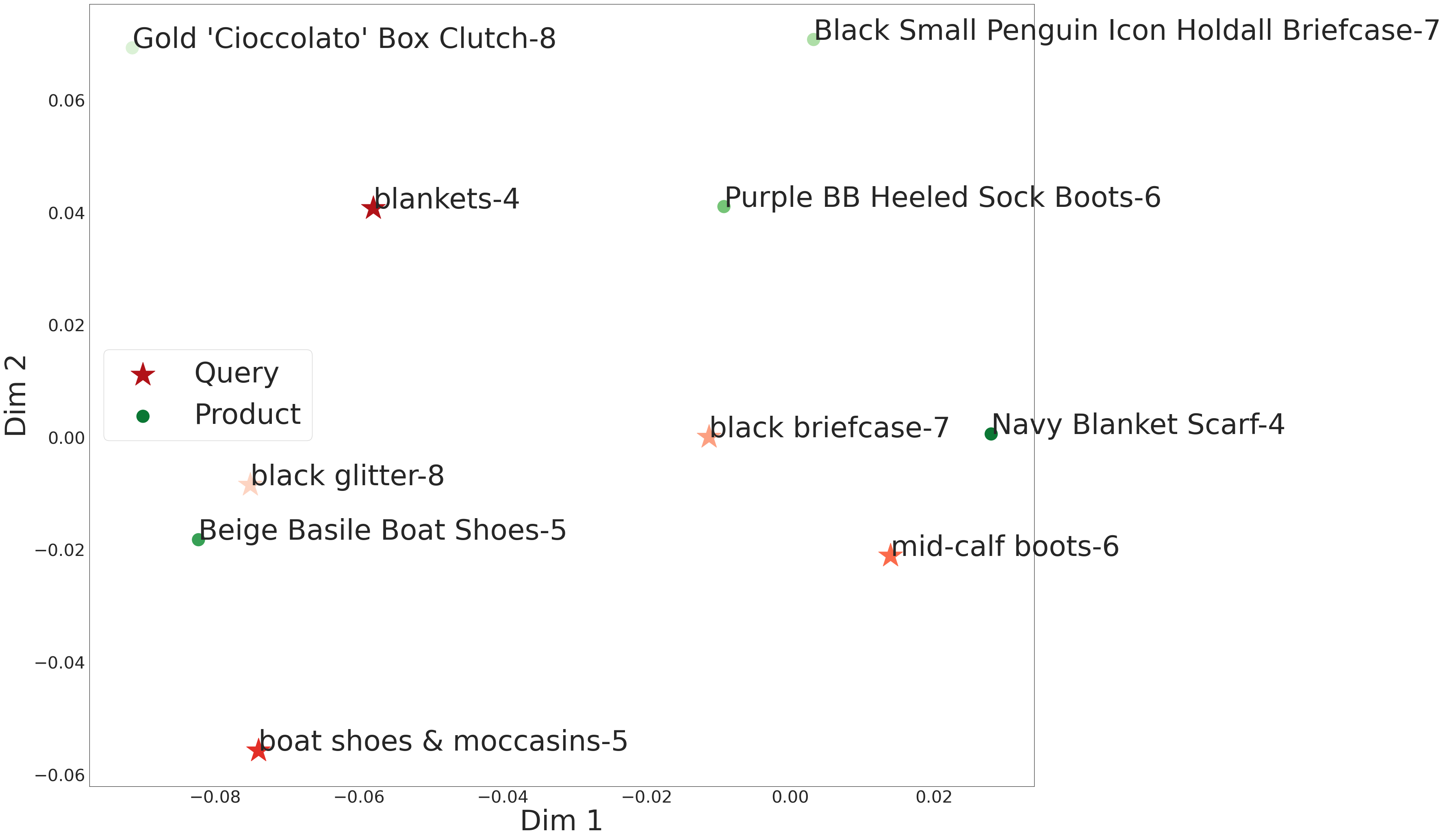}
  \caption{CLIP}
  \label{fig:tsne_clip}
\end{subfigure}
\begin{subfigure}{0.49\linewidth}
  \centering
  \includegraphics[width=\linewidth, trim={0 0 0 0}, clip]{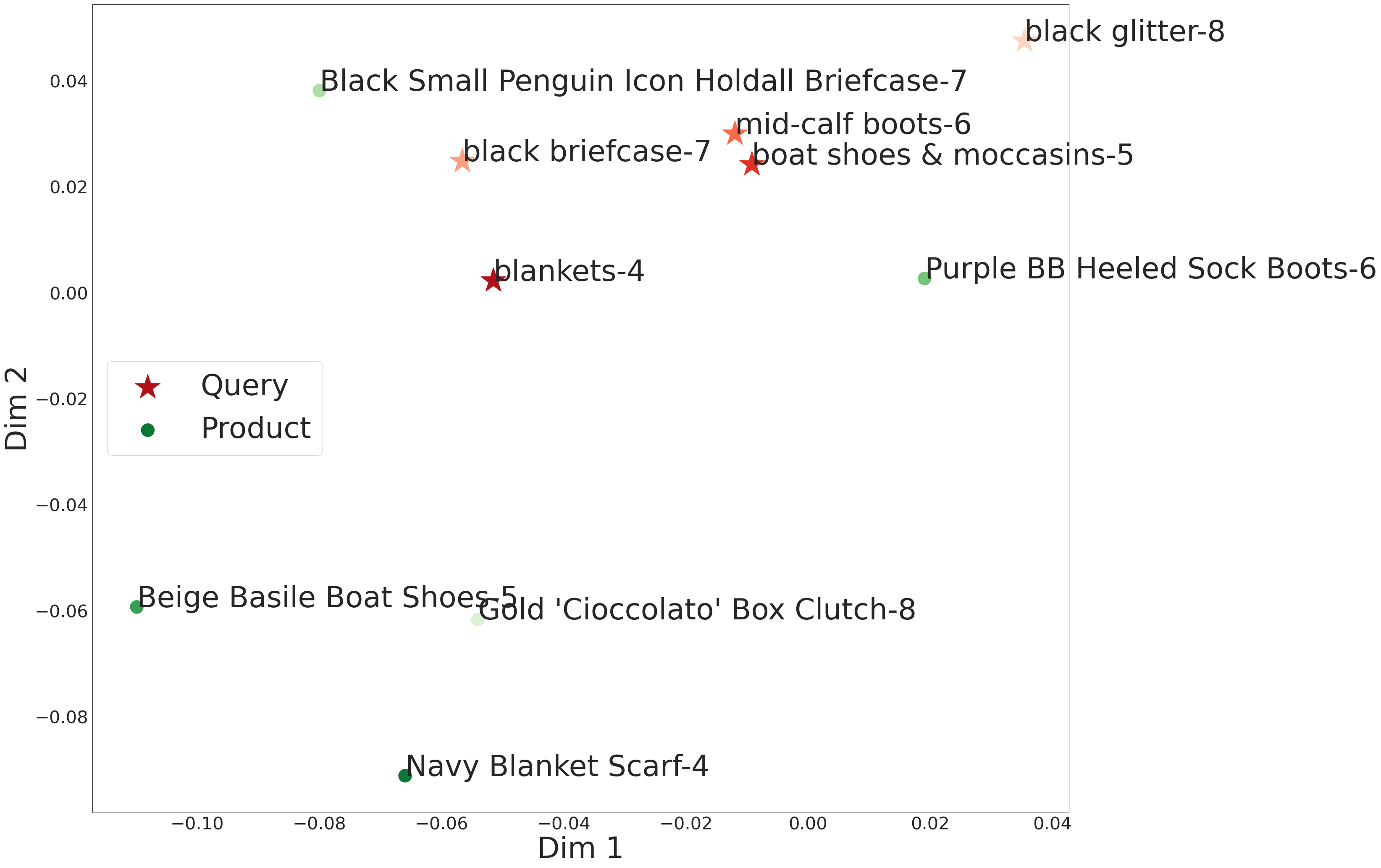}
  \caption{RetroMAE}
  \label{fig:tsne_retro}
\end{subfigure}
\begin{subfigure}{\linewidth}
  \centering
  \includegraphics[width=0.49\linewidth, trim={0 0 0 0}, clip]{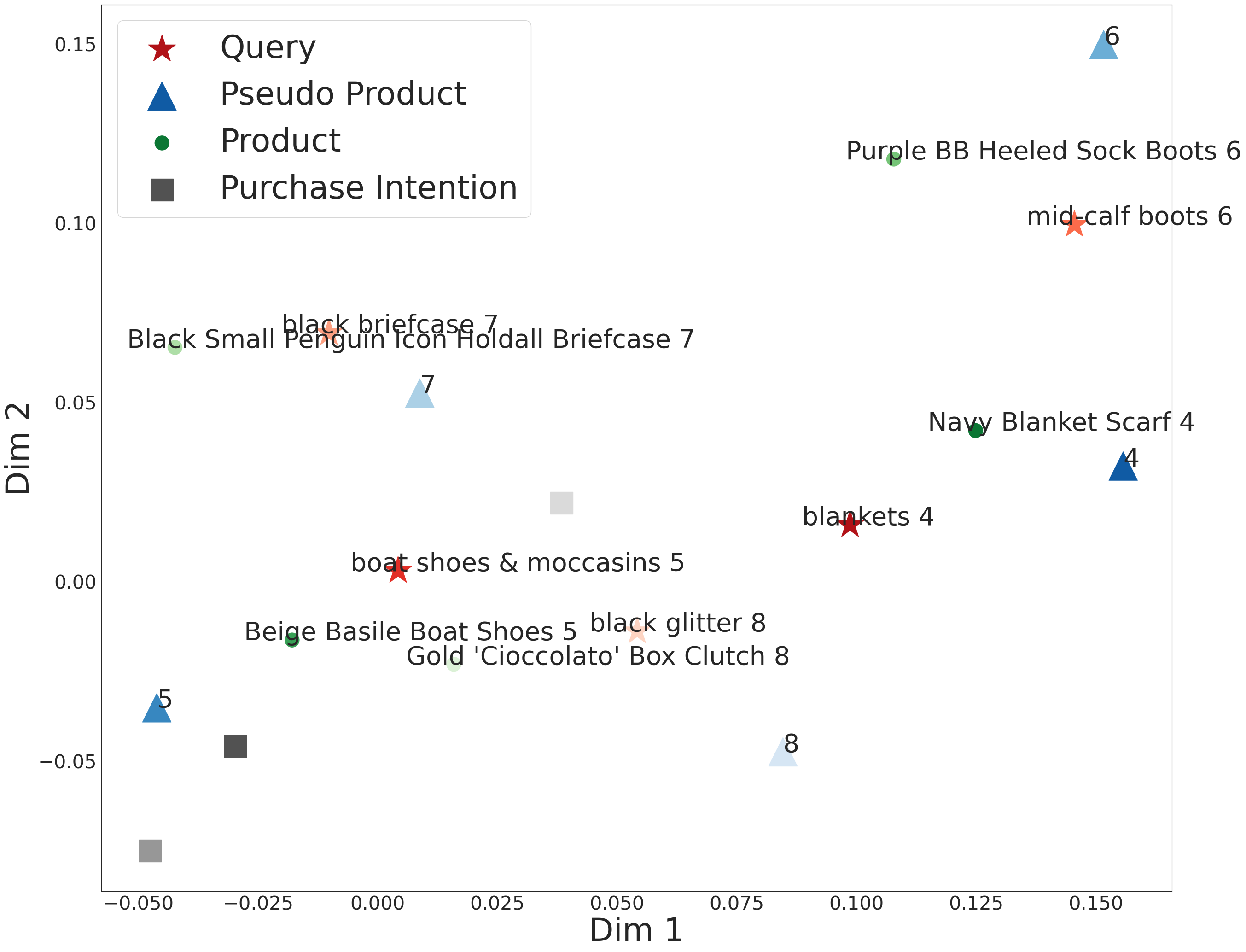}
  \caption{PINCER}
  \label{fig:tsne_pincer}
\end{subfigure}
\caption{t-SNE distribution of FashionGen(mean gradient) randomly chosen query-product pairs}
\label{fig:tsne_results}
\vspace{-15pt}
\end{figure*}

\begin{figure}[!h]
\centering
\includegraphics[width=\linewidth, trim={0cm 0 0cm 0}, clip]{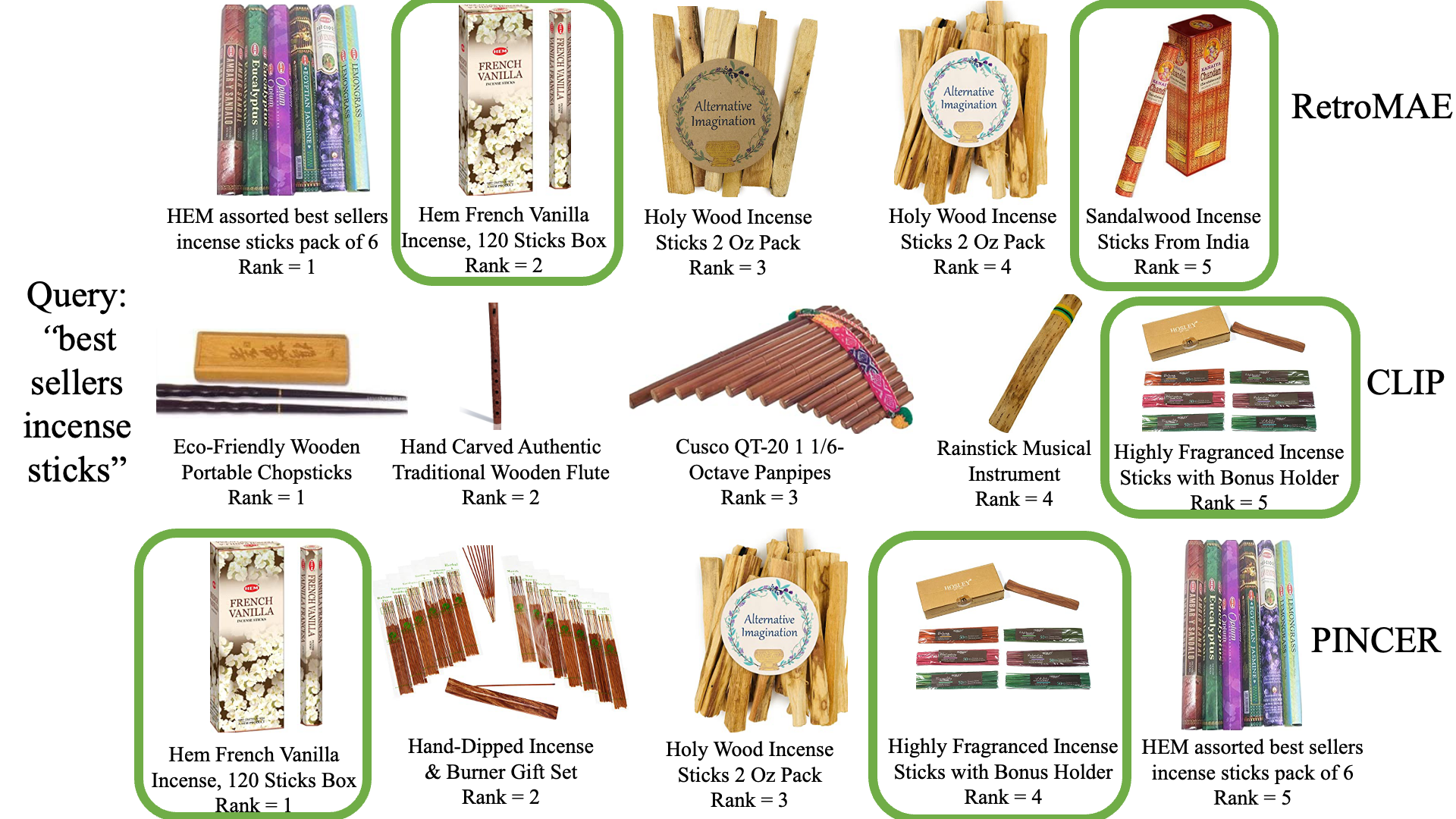}
\caption{Retrieval comparison between RetroMAE, CLIP, and PINCER models for a query from Amazon (brightness) test dataset}
\label{fig:am_image_results}
\vspace{-5pt}
\end{figure}

\begin{figure}[!h]
\centering
\includegraphics[width=\linewidth, trim={0cm 0 0cm 0}, clip]{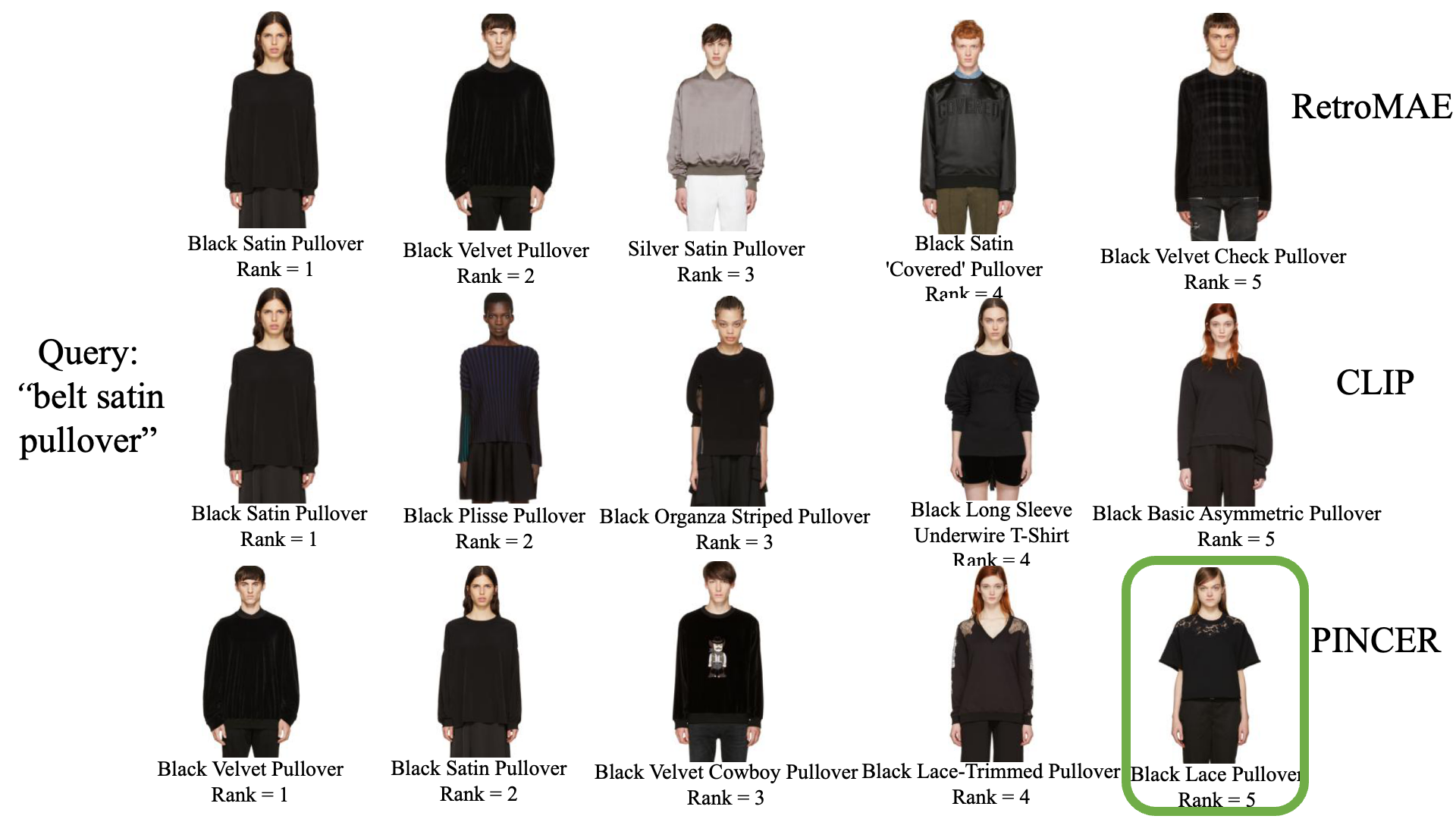}
\caption{Retrieval comparison between RetroMAE, CLIP, and PINCER models for a query from FashionGen (mean gradient) test dataset}
\label{fig:fg_image_results}
\vspace{-10pt}
\end{figure}

Fig. \ref{fig:tsne_results} provides close-up views of the t-SNE plots for CLIP, RetroMAE, and PINCER, revealing their distinct query-product pair distributions. CLIP (Fig. \ref{fig:tsne_clip}) exhibits distinct pockets of query-product pairs, demonstrating a strict one-to-one relationship that increases precision but reduces the ability to retrieve relevant products, decreasing recall. RetroMAE (Fig. \ref{fig:tsne_retro}) groups products in distinct neighborhoods, with queries positioned around them, contributing to good retrieval performance (Tables \ref{tab:real_pre_rec} \& \ref{tab:syn_pre_rec}). However, some products belonging to different neighborhoods are well separated, but the queries relevant to those products overlap with other neighborhoods, reducing recall. In contrast, PINCER (Fig. \ref{fig:tsne_pincer}) transforms queries into pseudo products by locating relevant purchase intention vectors, reducing the need for query rephrasing and improving the retrieval of relevant products. PINCER strategically distributes query and product embeddings around the purchase intention vectors, which act as grouping centers, clustering all pairs around their nearest vectors. This approach increases the model's ability to retrieve relevant products and improves recall while efficiently diversifying the retrieval process. The purchase intention vectors also highlight their potential use for real-time retrieval without requiring additional vector indexing libraries.

The figures \ref{fig:am_image_results}, and \ref{fig:fg_image_results} compare the retrieval results from RetroMAE, CLIP, and PINCER for various test queries, each associated with five purchased products. A green box around a retrieved product image indicates a match with a purchased product. For the query \textbf{\textit{"best sellers incense sticks"}} (Fig. \ref{fig:am_image_results}), RetroMAE and PINCER retrieve incense stick products that are possibly best-sellers, matching the purchased products, while CLIP only retrieves one relevant product. RetroMAE focuses more on text matching of \textbf{\textit{"best sellers"}} brand, whereas PINCER demonstrates its ability to capture the underlying purchase intention of most sold incense sticks including the brand. This results in more relevant and diverse product retrievals compared to the other models.

Fig. \ref{fig:fg_image_results} presents the retrieval results for the query \textbf{\textit{"belt satin pullover"}}. Among the three models, only PINCER successfully retrieved a matching product within the top five results. Although all the models retrieved pullover products, PINCER's results specifically included a product that was part of the purchased products list. This can be attributed to the preference modeling of purchased products, where the pseudo product embeddings are drawn closer to the user's potentially purchasable products. This phenomenon makes PINCER superior to other models in terms of achieving high precision \& recall in product retrieval tasks.

\section{Ablation Study}
\begin{table}[ht]
\fontsize{9pt}{9pt}\selectfont
\centering
\begin{tabular}{|c|p{1cm}p{1cm}p{1cm}p{1cm}|c|}
\hline
\multirow{2}{*}{PINCER Model} & \multicolumn{4}{c|}{Recall}                                                               & \multirow{2}{*}{SumR} \\ \cline{2-5}
                       & \multicolumn{1}{p{0.25cm}|}{R@10} & \multicolumn{1}{p{0.25cm}|}{R@20} & \multicolumn{1}{p{0.25cm}|}{R@50} & R@100 &                       \\ \hline
\begin{tabular}[c]{@{}c@{}}Stage 1:\\only PF\end{tabular}   & \multicolumn{1}{c|}{27.31} & \multicolumn{1}{c|}{40.54} & \multicolumn{1}{c|}{60.17} & 72.78 & 201 \\ \hline
\begin{tabular}[c]{@{}c@{}}Stage 1:\\only PI\end{tabular} & \multicolumn{1}{c|}{27.49} & \multicolumn{1}{c|}{40.78} & \multicolumn{1}{c|}{60.32} & 73.14 & 202 \\ \hline
\begin{tabular}[c]{@{}c@{}}Stage 1 + Stage 2:\\PI+PF\end{tabular}             & \multicolumn{1}{c|}{31.85} & \multicolumn{1}{c|}{46.36} & \multicolumn{1}{c|}{67.08} & 79.01 & 224 \\ \hline
\end{tabular}
\vspace{8pt}
\caption{PINCER ablation study from FashionGen (mean gradient) dataset}
\label{tab:abl_study}
\vspace{-10pt}
\end{table}

The results presented in Table \ref{tab:abl_study} underscore the performance gains achieved by PINCER with and without the integration of Purchase Intention (PI) and Product Features (PF). This ablation study uses one of the synthetic datasets to showcase the model functionality because the prior experimental results prove the validity of the datasets. In Table \ref{tab:abl_study}, Stage 1 uses PI and PF individually to show their individual contribution to the model performance. The ablation study conducted affirms the pivotal role played by purchase intention and product features, as their inclusion significantly enhances the training outcomes in Stage 2. Without PI and PF, PINCER's two-stage training process would essentially resemble a two-tower CLIP architecture. It is important to note that PI and PF are integral components of PINCER that influence the stability of Stage 2 training. The stage 2 cannot be evaluated in isolation with only one of the two components present. Nonetheless, the pseudo product embeddings generated by PINCER demonstrate the criticality of utilizing purchase intention and product features, as it outperforms existing multi-modal and text retrieval models.

\section{Real-time retrieval Performance}
\begin{figure}[ht]
\centering
\begin{subfigure}{0.49\linewidth}
  \centering
  \includegraphics[scale=0.145]{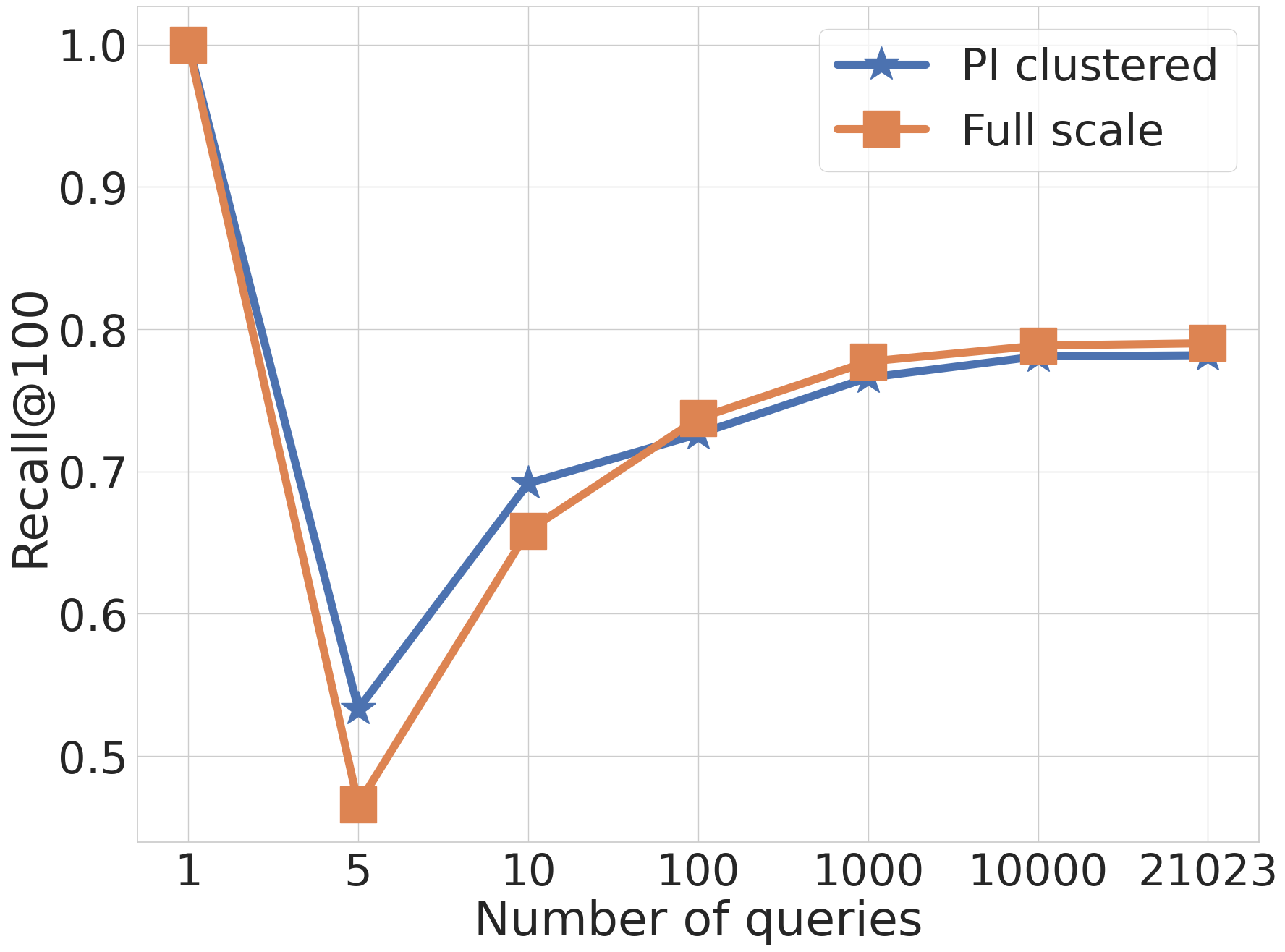}
  \caption{Retrieval performance}
  \label{fig:recall_retrieval}
\end{subfigure}%
\begin{subfigure}{0.49\linewidth}
  \centering
  \includegraphics[scale=0.145]{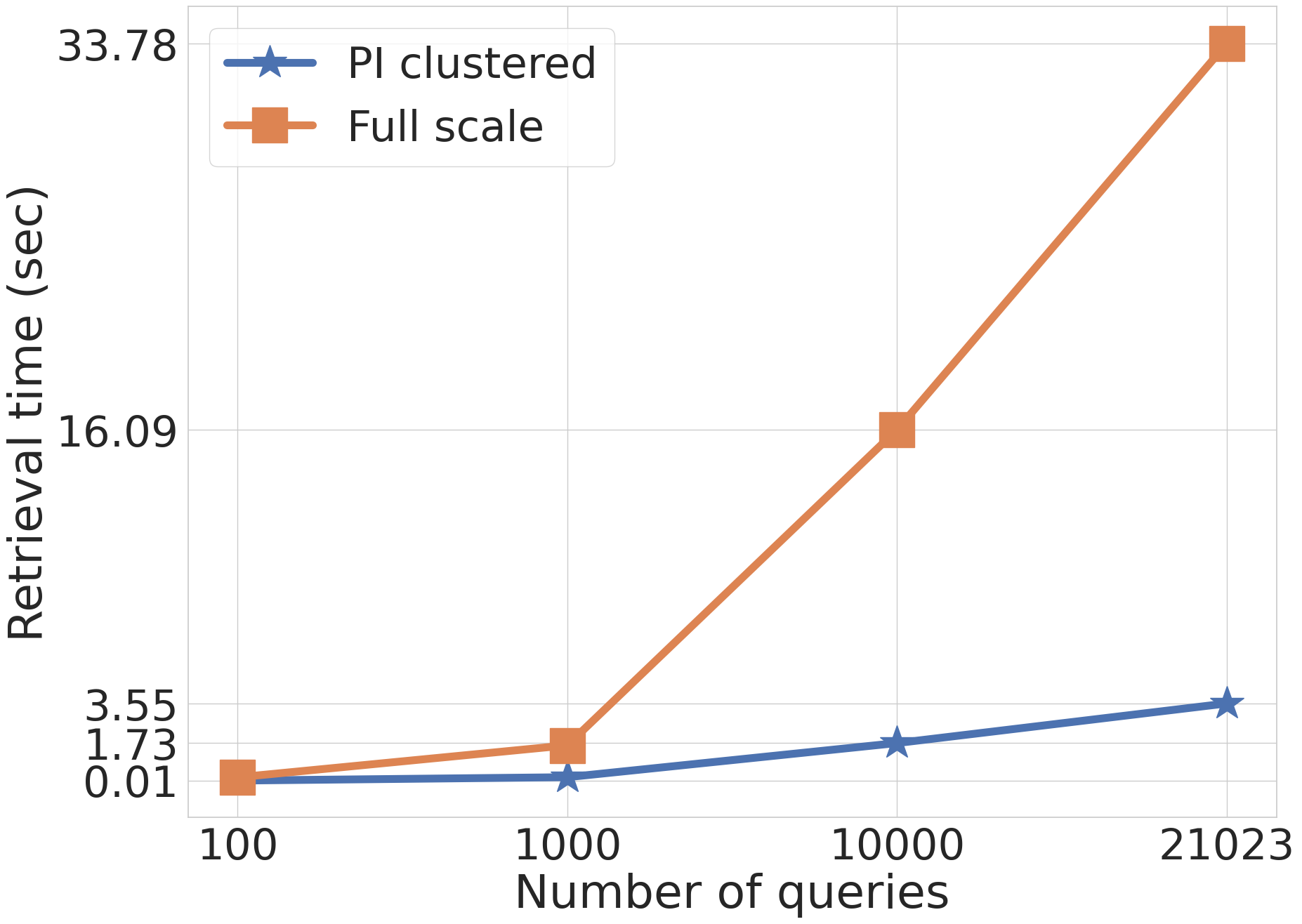}
  \caption{Retrieval time}
  \label{fig:time_retrieval}
\end{subfigure}
\caption{Retrieval comparison between full-scale and PINCER clustered products}
\label{fig:retrieval-results}
\vspace{-10pt}
\end{figure}
Fig. \ref{fig:retrieval-results} illustrates the comparison between full-scale retrieval and Purchase Intention (PI) clustered products from the PINCER algorithm, encompassing $21023$ queries and $67$K unique products. Post-training, we retain all product embeddings and apply nearest neighbor algorithm for each cluster using PI vectors as cluster centroids. The use of cluster centroid indices from PI embeddings enables the matching of incoming queries with relevant cluster groups. While Fig. \ref{fig:recall_retrieval} highlights a slight degradation in PINCER's performance at $100$ queries and beyond with clustered retrieval, Fig. \ref{fig:time_retrieval} demonstrates its advantage in retrieval time. The observed drop in $Recall@100$ can be attributed to the specificity constraints of product clustering within the purchase intent vectors. As queries retrieve from a narrower cluster subset, fewer highly relevant matches may be found in the larger recall range, thereby impacting the overall $Recall@100$ score. Clustered retrieval for top@100 on $1000$ queries achieves a $15$ms latency compared to the $30$ms real-time retrieval requirement. This comparison between full-scale and clustered retrieval showcases PINCER's potential for deployment in real-time retrieval platforms within a scalable environment.

\section{Limitations}
PINCER's reliance on contrastive learning makes its performance scale with batch size and GPU power. This study is limited to very few open-source models that are trainable with in the available GPU power.  

\section{Conclusion}
In conclusion, this work introduces PINCER, a novel multi-modal transformer framework that bridges the gap between queries and purchased products in e-commerce retrieval by modeling potential user purchase intention. Through a two-stage training process, PINCER estimates purchase intention via reward-based competitive learning, stores and retrieves granular product features, optimizes relevance ranking, and generates pseudo product embeddings close to the target. Experimental results highlight PINCER, an efficient framework that outperform existing retrieval models by capturing user purchase nuances. PINCER advances e-commerce retrieval techniques in a systematic process to improve the retrieval recall. Future directions involve enabling PINCER to accept multi-modal query input, including user profile information from click-stream data with purchase intention for personalized search and product recommendations.

\printbibliography


\appendix

\section{Appendix / supplemental material}
\subsection{\textbf{Algorithm:}}
The query transformation contains pre-trained language model encoder, latent embedding projectors, purchase intention vectors, and decoder. The product encoding uses pre-trained language and vision model encoders and latent embedding projects. The framework employ the pre-trained models to leverage the world knowledge of text and images to generate semantically relevant embeddings in a shared latent space. 

The framework efficiently train the query and the product modules in stage 1 to align the purchase intention vectors with the query and product embeddings, and query-product granular features. Stage 2 training uses the same stage 1 data to generate the pseudo-product embedding. The purchase intentions are a fixed number of vectors that are shared between the queries and products, and instigate a shared learning.

A competitive learning strategy trains purchase intentions, rewarding encoders when query and product choose the same intent vector.

The training stages in Fig. \ref{fig:PINCER_arch} use the following algorithms for training.

\RestyleAlgo{ruled}
\SetAlgoLined
\LinesNumbered
\SetKwComment{Comment}{\# }{}
\SetKwInput{kwInit}{Initialize}
\begin{algorithm}[ht]
\caption{PINCER Stage 1 Training}
    \KwIn{Query-product training pairs $(x_n, y_n) \in \mathbb{T}$ (batches) $\subset \mathbb{D}$(training data), learning rate $\alpha$, temperature $T$, batch size $B$, number of purchase intention vectors $|S|$}
    \kwInit{Pre-trained query encoder $E_q$, product text $E_{p_{t}}$ and image encoder $E_{p_{i}}$, query text $P_{q_{t}}$ and image $P_{q_{i}}$ projectors, product text $P_{p_{t}}$ and image $P_{p_{i}}$ projectors, purchase intention vectors $S$}
    \KwOut{Trained PINCER $E_q$, $E_{p_{t}}$, $E_{p_{i}}$, $P_{q_{t}}$, $P_{q_{i}}$, $P_{p_{t}}$, $P_{p_{i}}$, and $S$}
    \For{$epoch = 1$ to $max\_epochs$}{
        \For{$batch = 1$ to $\frac{|\mathbb{T}|}{B}$}{
            batch $\gets \{(x_n, y_n)\}_{n=1}^B$\\
            Encode queries $x_{n} \gets (x_{n_{t}}, x_{n_{i}})$ and products $y_{n} \gets (y_{n_{t}}, y_{n_{i}})$\\
            \For{$n = 1$ to $B$}{
                Calculate $s_{k_x}$, and $s_{k_y}$ from (\ref{eq:nearest_intent})\\
                $r_{s_k}$, $p_{x_n, s_{k_x}}$, $p_{y_n, s_{k_y}}$, $rp_{x_n, s_{k_x}}$, and $rp_{y_n, s_{k_y}}$ from (\ref{eq:reward}), (\ref{eq:prob}), and (\ref{eq:remaining_prob}) respectively \\
                }
            Calculate $L_{qpt}$, $L_{qpi}$, and $L_{qp}$ using contrastive loss from (\ref{eq:contrastive_loss})\\
            RCL Loss from (\ref{eq:pur_int_rll})\\
            Stage 1 loss from (\ref{eq:stage_1_loss})\\
            Update $E_q$, $E_{p_{t}}$, $E_{p_{i}}$, $P_{q_{t}}$, $P_{q_{i}}$, $P_{p_{t}}$, $P_{p_{i}}$, and $S$ using $L_{stage1}$ and learning rate $\alpha$
            }
        }
    \algorithmicreturn\ $E_q$, $E_{p_{t}}$, $E_{p_{i}}$, $P_{q_{t}}$, $P_{q_{i}}$, $P_{p_{t}}$, $P_{p_{i}}$, and $S$\\
\label{alg:stage_1}
\end{algorithm}

\begin{algorithm}[ht]
\SetAlgoLined
\caption{PINCER Stage 2 Training}
    \KwIn{Query-product training pairs $(x_n, y_n) \in \mathbb{T}$, learning rate $\alpha$, temperature $T$, batch size $B$, stage 1 trained encoders ($E_q$, $E_{p_{t}}$, $E_{p_{i}}$), projectors ($P_{q_{t}}$, $P_{q_{i}}$, $P_{p_{t}}$, $P_{p_{i}}$), product features $F$, and purchase intention vectors $S$}
    \kwInit{Randomly initialize learnable vector $L_{v}$, transformer decoder $D$}
    \KwOut{Trained PINCER decoder $D$, and learned vector $L_{v}$}
    \For{$epoch = 1$ to $max\_epochs$}{
        \For{$batch = 1$ to $\frac{|\mathbb{T}|}{B}$}{
            batch $\gets \{(x_n, y_n)\}_{n=1}^B$\\
            Encode queries $x_{n} \gets (x_{n_{t}}, x_{n_{i}})$ and products $y_{n} \gets (y_{n_{t}}, y_{n_{i}})$\\
            \For{$n = 1$ to $B$}{
                $s_{k_n} \gets argmin_{s \in S} \|x_n - s\|_2$ \Comment{Nearest intent for $x_n$}
                $f_{j_n} \gets argmax_{f \in F} \cos(x_n, f)$ \Comment{$x_n$ relevant product features}
                $\hat{y}_n \gets D(x_n, s_{k_n}, f_{j_n})$ \Comment{Generate pseudo product}
            }
            Calculate $PML$ from (\ref{eq:pref_reward})\\
            Calculate $SD_{\hat{y}_n, y_n}$ and $SD_{x_n, y_n}$ from (\ref{eq:sim})\\
            Update $D$, and $L_{v}$ using $L_{stage2}$ and learning rate $\alpha$\
        }
    }
    \algorithmicreturn\ {$D$, and $L_{v}$}\\
\label{alg:stage_2}
\end{algorithm} 

\begin{figure*}[!h]
  \centering 
  \includegraphics[width=\linewidth]{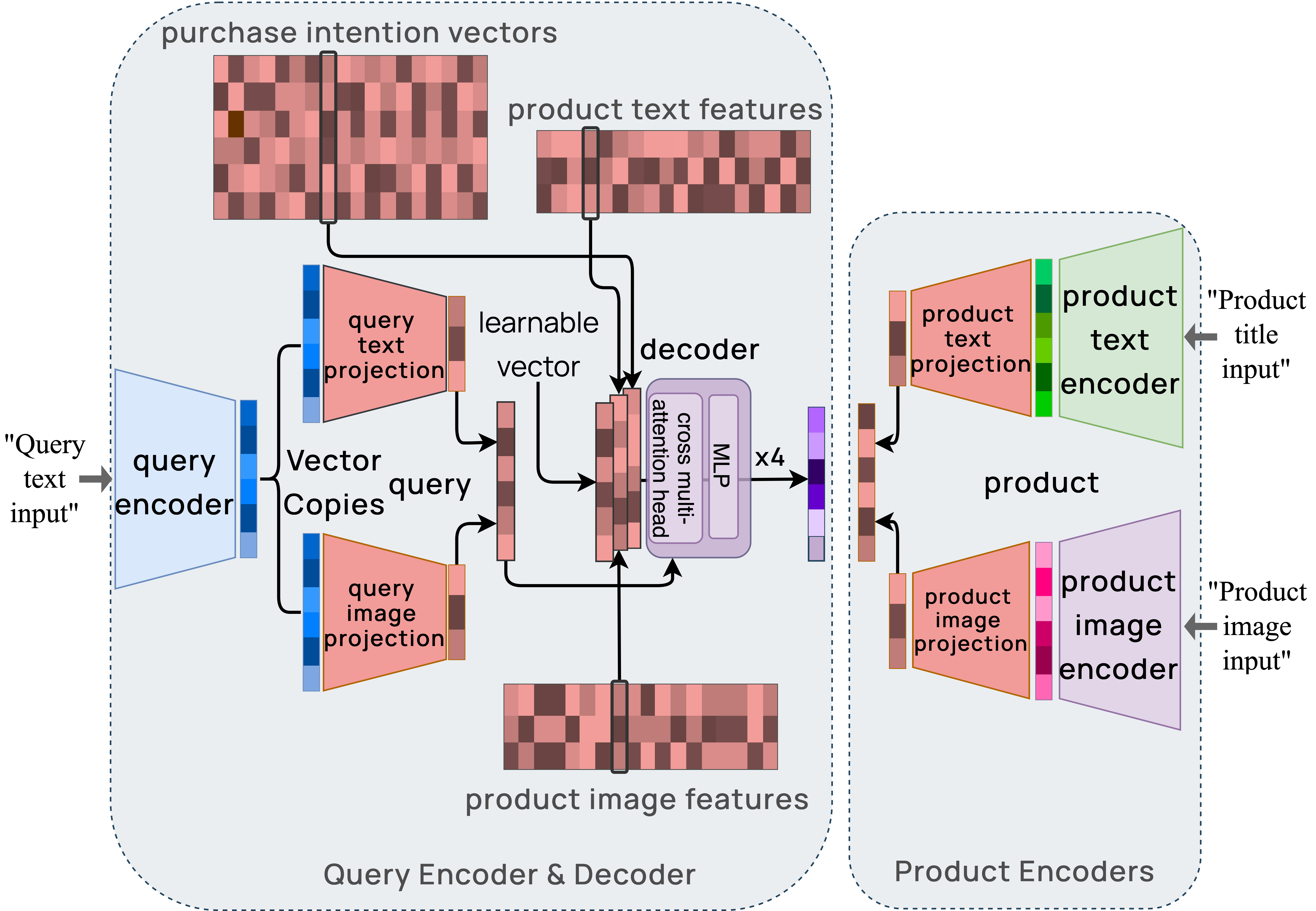} 
  \caption{PINCER model architecture} 
  \label{fig:PINCER_full_model} 
\end{figure*}

\end{document}